# Electronic Properties and Bonding in ZrH$_x$ Thin Films Investigated by Valence-Band X-ray Photoelectron Spectroscopy

Martin Magnuson, Susann Schmidt, Lars Hultman, and Hans Högberg

*Thin Film Physics Division, Department of Physics, Chemistry and Biology (IFM), Linköping University, SE-58183 Linköping, Sweden*

**Abstract**

The electronic structure and chemical bonding in reactively magnetron sputtered ZrH$_x$ ($x$=0.15, 0.30, 1.16) thin films with oxygen content as low as 0.2 at% are investigated by *4d* valence band, shallow *4p* core-level and *3d* core-level X-ray photoelectron spectroscopy. With increasing hydrogen content, we observe significant reduction of the *4d* valence states close to the Fermi level as a result of redistribution of intensity towards the H *1s* – Zr *4d* hybridization region at ~6 eV below the Fermi level. For low hydrogen content ($x$=0.15, 0.30), the films consist of a superposition of hexagonal closest packed metal (α-phase) and understoichiometric δ-ZrH$_x$ (CaF$_2$-type structure) phases, while for $x$=1.16, the film form single phase ZrH$_x$ that largely resembles that of stoichiometric δ-ZrH$_2$ phase. We show that the cubic δ-ZrH$_x$ phase is metastable as thin film up to $x$=1.16 while for higher H-contents, the structure is predicted to be tetragonally distorted. For the investigated ZrH$_{1.16}$ film, we find chemical shifts of 0.68 and 0.51 eV towards higher binding energies for the Zr *4p$_{3/2}$* and *3d$_{5/2}$* peak positions, respectively. Compared to the Zr metal binding energies of 27.26 and 178.87 eV, this signifies a charge-transfer from Zr to H atoms. The change in the electronic structure, spectral line shapes, and chemical shifts as function of hydrogen content is discussed in relation to the charge-transfer from Zr to H that affects the conductivity by charge redistribution in the valence band.

## 1. Introduction

Zirconium-hydrides are important in many applications in the nuclear industry [1], in getter vacuum pumps, as hydrogen storage, in powder metallurgy, and as zircaloy [2]. The hardness, ductility and tensile strength of ZrH$_x$ alloys can be controlled by varying the hydrogen content [3]. Increasing the H content results in substoichiometric δ-ZrH$_x$ phase (CaF$_2$-type structure, $x$ = ~1.6-2 [4] as shown in Fig. 1) and a body-centred tetragonal ε-phase ($x$ = 1.75–2) with a ThH$_2$-type structure [5] has also been observed [4]. The δ-ZrH$_x$ structure originates from incerting H atoms to occupy all or parts of the tetrahedral interstitials in the CaF$_2$ structure. Hydrogen acts as a hardening element that prevents dislocation movements and the hydride material becomes a ceramic that is harder, but less ductile than Zr metal [6].

Traditionally, bulk Zr-hydrides are synthesized by annealing Zr metal in hydrogen gas for a time period of days to a few weeks at high temperatures between 400-900 °C for a homogeneous diffusion process at different H-contents [7]. However, defects in the bulk of the Zr metal results in variations in the diffusion rate for hydrogen that cause ZrH$_x$ alloys with composition gradients and less well-defined polycrystalline structure including grain boundaries. At room temperature in air, ZrH$_x$ quickly form a nm thin surface oxide layer that prevents further oxidation into the bulk. With increased annealing temperature, oxygen proceeds deeper into the bulk of the material, in particular along grain boundaries [8], and a few percent of oxygen cannot be avoided with the annealing-diffusion hydration synthesis method. Consequently, the above conditions hinder adequate bonding determination in hydrides by spectroscopy.

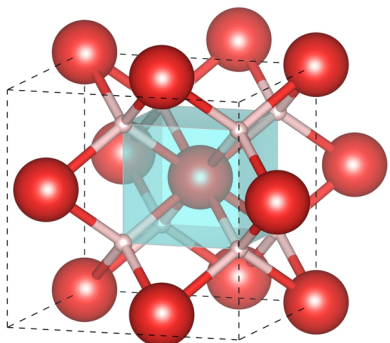

Figure 1: (Color online) Unit cell of stoichiometric δ-ZrH$_2$ (space group 225) with Zr-atoms (red spheres) in (0,0,0), (½,½,0), (½,0,½), and (0,½,½) and with H (white spheres) in all tetrahedral sites (¾,¼,¼), (¼,¾,¼), (¼,¼,¾), (¾,¾,¾), (¼,¼,¼), (¾,¾,¼), (¼,¾,¾), and (¾,¼,¼).

Previous experiments on bulk ZrH$_x$ using ultraviolet photoelectron spectroscopy showed significant changes in the electronic structure for high hydrogen contents ($x$=1.63-1.94) both at the states near the Fermi level (E$_F$) and in deeper-lying states around -7 eV [9]. These changes were associated with a phase transition from a cubic (fcc) structure to a tetragonal (fct) structure as a result of spontaneous symmetry breaking that removes the degeneracy by a distortion known as the Jahn-Teller effect [10] that lowers the total energy of the $4d^2 5s^2$ electron configuration in the Zr valence band.

For high hydrogen contents ($x$=1.63-1.9), core-level X-ray photoelectron spectroscopy (XPS) studies of bulk ZrH$_x$ materials indicated a significant shift of the Zr $4p$ and $3d$ core-levels by 1.0 eV and 0.7 eV [7], respectively. For $x$=1.52-1.68, valence band XPS also showed an interstitial Zr-H bonding peak around 6.4 eV below E$_F$ [11]. Although these studies indicate significant changes in the Zr-H bond for different hydrogen contents, the spectra are affected by superimposed O $2p$ states that occur in the same energy region (5-8 eV) as the Zr-H bonding structures [12].

For the ZrH$_x$ thin films in this work, we apply XPS at the Zr $3d$, $4p$ core-levels, and the $4d$ valence band to investigate the electronic structures to determine the chemical bonding and conductivity properties as a function of relatively low hydrogen content ($x$=0.15, 0.30, and 1.16) and compare the result to bulk α-Zr as reference. These hydride compositions were chosen to obtain a spread in the type and amount of Zr-H bonds. Deposited δ-ZrH$_x$ films with CaF$_2$-type structure, are found to be stable outside the homogeneity region determined for bulk δ-ZrH$_2$ with compositions ranging from about ZrH$_{1.5}$ to stoichiometric [13]. To control the structure and hence the electronic properties of the materials and to overcome the problems with oxidation, we deposited homogeneous thin films of ZrH$_x$ using reactive direct current magnetron sputtering (rDCMS). The thin films were grown without external heating and contamination of oxygen and other contaminates were very low. Thin films are particularly favorable for correct characterization of the material's fundamental bonding properties and to identify different phases in understoichiometric transition metal hydrides.

## 2. Experimental
### 2.1 Thin film deposition and XRD characterization

The Zr-H films studied were deposited on Si(100) substrates by rDCMS. ZrH$_x$ films with $x$=0.15, 0.30, and 1.16 were deposited in an industrial high vacuum coating system (CemeCon AG, Würselen, Germany). Here, a zirconium target was sputtered in Ar (99.9997%)/H$_2$ (99.9996%) mixtures using a fixed Ar partial pressure of 0.42 Pa with 5, 10, and 20% H$_2$. The substrate bias was set to -80 V and no external substrate heating was applied. The films were deposited for 120 s resulting in films with thicknesses ranging between ~800 to ~840 nm. More detailed information on the deposition conditions are presented in ref [14]. The investigated α-Zr reference sample was a commercial zirconium target with a purity of 99.9% from (Kurt J. Lesker Company, Clairton, PA, USA).

## 2.2 XPS measurements

Valence band (VB) and XPS measurements of the Zr *3d*, *4p*, *4d* valence band and O *1s* core-level regions were performed in a surface analysis system (AXIS UltraDLD, Kratos, Manchester, U.K.) using monochromatic Al-$K_\alpha$ (1486.6 eV) radiation with an incidence angle of 54° and a spot size of 300 x 800 μm. The electron energy analyzer detected photoelectrons perpendicular to the sample surface with an acceptance angle of ±15°. The spectra were recorded with a step size of 0.1 eV and a pass energy of 10 eV, which provided an overall energy resolution better than 0.5 eV. The binding energy scale of all XPS spectra was referenced to the $E_F$, which was set to a binding energy of 0 eV [15]. The $ZrH_x$ thin films in this work were examined before and after $Ar^+$ sputtering for 600 s using an $Ar^+$ incident angle of 20º at 4 keV rastered over an area of 2x2 $mm^2$.

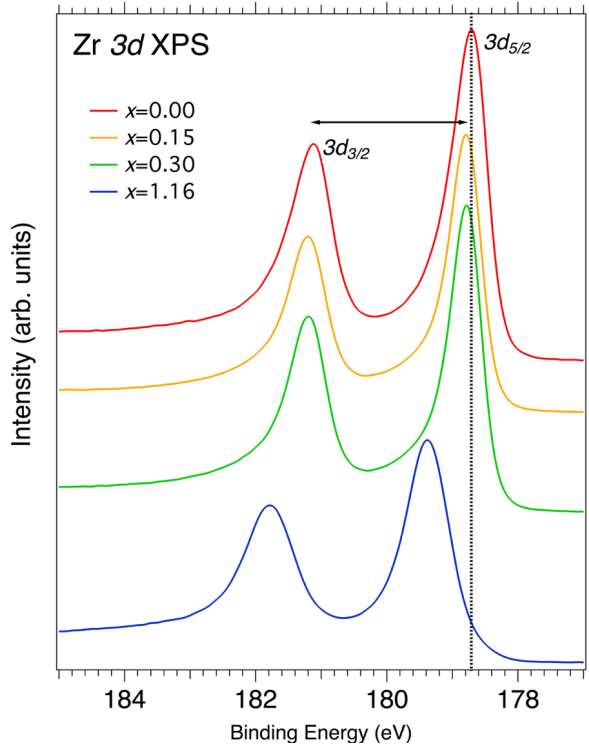

**Figure 2:** Zr *3d* core-level XPS spectra of the $ZrH_x$ thin films in comparison to α-Zr metal. The *3d$_{5/2,3/2}$* spin-orbit splitting (2.41 eV) is indicated by the horizontal arrow.

## 2.3 Structural model and DFT Calculations

The geometry relaxation was performed using the Perdue-Burke-Ernzerhof (PBE) functional [16] including the Grimme van der Waals density functional theory (DFT)-D2 scheme [17]. The first principle calculations were carried out using density functional theory (DFT) implemented in the Vienna *ab initio* simulation package (VASP) [18] with an exchange-potential functional using the General Gradient Approximation (GGA). A hydrogen-potential suited for short bonds (hydrogen PAW potential H_h) with an energy cut-off of 1050 eV was used. For the self-consistent calculations, the projected augmented wave (PAW) [19] method was used with the PBE and the Heyd-Scuseria-Ernzerhof functionals of the generalized gradient approximation. The HSE06 functional [20] is a hybrid functional including a linear combination of short and long-range PBE exchange terms and a short-range Hartree-Fock term that improves the formation energies and band gaps and peak positions. The screening parameter of HSE06 was 0.2 $Å^{-1}$ with a plane wave cutoff energy of 700 eV. A 29x29x29 *k*-grid was used for the standard PBE functional for the structure relaxation and a 11x11x1 *k*-grid was used for HSE06 functional as the computational cost is significantly higher for the HSE06 in comparison to the PBE functional. Using the HSE06 exchange correlation functional, the energy positions of the bands are shifted by ~1.8 eV towards higher energy relative to the $E_F$, and are thus much more realistic, in comparison to the corresponding energy positions, using the PBE functional. The charge-transfer calculations between the different elements were made using standard BADER analysis [21].

## 3. Results and Discussion

Figure 2 shows Zr *3d* core-level XPS spectra of the ZrH$_x$ thin films in comparison to α-Zr metal. As observed, there is a significant chemical shift towards higher binding energies for the film grown with hydrogen in the plasma compared to the pure metal film, which is a consequence of the higher electronegativity of H (χ =2.2) in comparison to Zr (χ=1.33) [22]. For *x*=1.16, the Zr *3d$_{5/2}$* peak position was determined to 179.38 eV. This is in good agreement with the Zr *3d$_{5/2}$* binding energy of 179.375 reported for bulk δ-ZrH$_{1.74}$ [23] as well as 179 eV determined for bulk ZrH$_{1.9}$ material [24] and close to the 180 eV binding energy determined for bulk ZrH$_{1.64}$ [25] and ZrH$_{1.9}$ [7]. Furthermore, the position of the *3d$_{5/2}$* peak is more or less identical to the binding energy determined for the reactively sputtered ZrH$_{1.64}$ film in ref. 14 with 179.44 eV. Note the small energy shift towards lower binding energy for the less hydrogen rich film with *x*=1.16 (179.38 eV) that is consistent with a smaller charge-transfer from Zr to H in this film in comparison to the *x*=1.64 film (179.44 eV). In addition, for the ZrH$_{1.16}$ film, there is a high-energy shift of +0.51 eV while between *x*=0.15 and 0.30, the Zr *3d* core-level high-energy shift is much smaller (+0.06 eV) and is an indication of smaller average charge-transfer from Zr towards H in these samples.

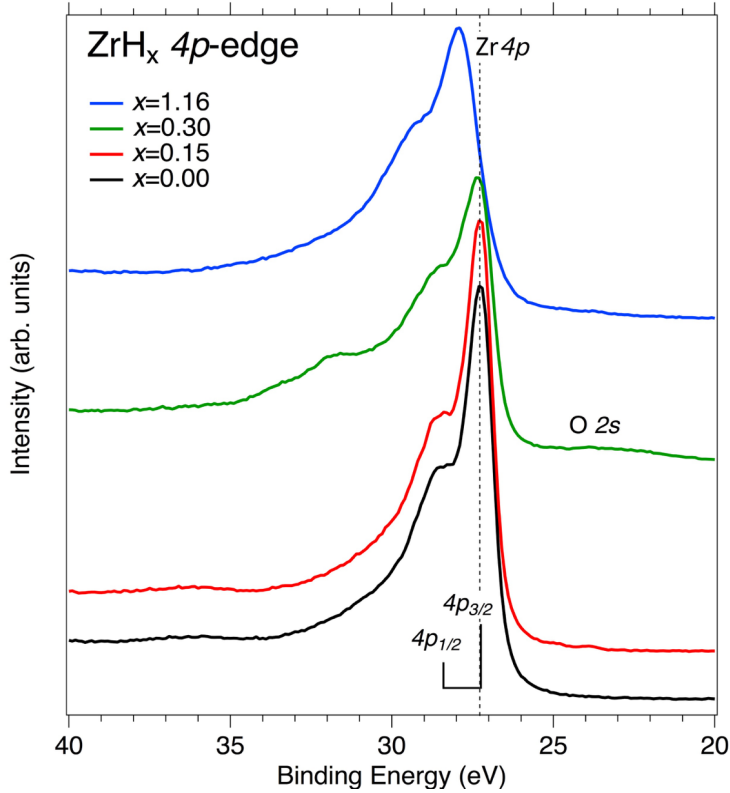

**Figure 3:** Zr *4p* core-level XPS spectra of the ZrH$_x$ thin films in comparison to α-Zr metal. The *4p$_{3/2,1/2}$* spin-orbit splitting (1.3 eV) is indicated by the vertical lines, from ref. [26].

The metallic reference sample has the Zr *3d$_{5/2}$* and Zr *3d$_{3/2}$* peak positions at 178.87 and 181.28 eV (2.41 eV peak splitting) in good agreement with literature values [26]. The Zr *3d$_{5/2}$* - *3d$_{3/2}$* spin-orbit splitting is the same in the films. Prior to sputtering, the spectra showed structures of ZrO$_2$ following air exposure, but at the Zr *3d*-edge after sputtering there are no features connected to oxygen.

To evaluate the asymmetric tails towards higher binding energies in each spectrum, a Doniac-Sunjic function [27], corresponding to the electron-hole pair excitations created at the Fermi level to screen the core hole potential was applied. The Doniac-Sunjic profile is essentially a convolution between a Lorentzian with the function 1/E(1-α), where E is the binding energy of each peak and α is a parameter known as the *singularity index* that is related to the electron density of states at the Fermi level [28]. The intensities of the tails are related to the amount of metallicity in the system due to the coupling of the core-hole with collective electron oscillations. For α-Zr, the singularity index is large (α=0.15) and the *3d$_{5/2}$*/*3d$_{3/2}$* branching ratio is higher (1.75) than the statistical value of 1.67 (5/3), signifying high conductivity. For *x*=0.15 and 0.30, the singularity index and branching ratio remains almost the same as for α-Zr, but for $x$=1.16, the tail becomes significantly smaller with a singularity index of only α=0.04 with a branching ratio that is smaller (1.61) than the statistical ratio (1.67). This shows that the number of bands crossing the $E_F$ are reduced and thus the expected conductivity.

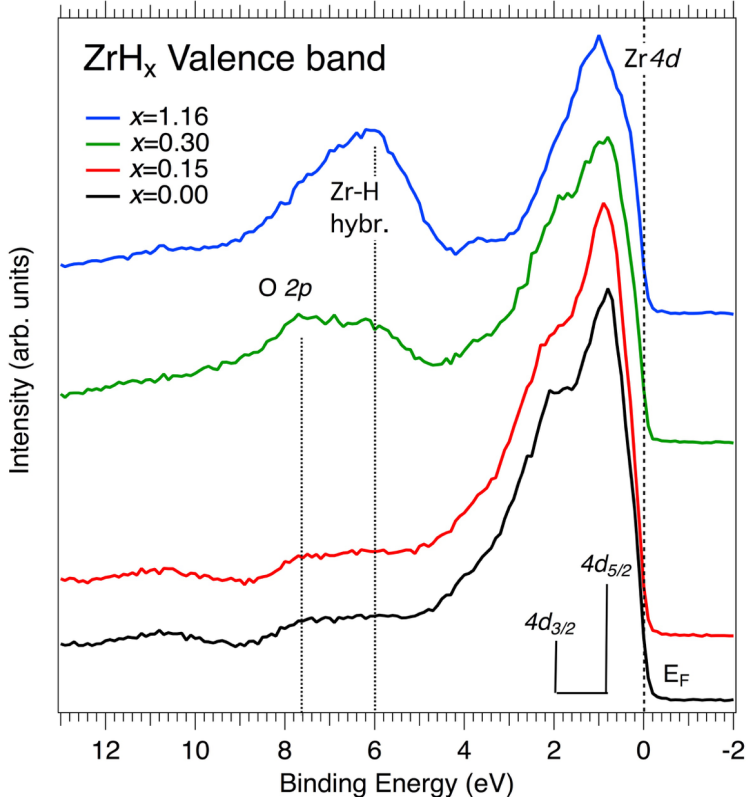

**Figure 4:** Valence band XPS spectra of the ZrH$_x$ thin films in comparison to α-Zr metal.

Figure 3 shows Zr *4p* shallow core-level XPS spectra of the ZrH$_x$ thin films in comparison to α-Zr metal. As in the case of the Zr *3d* level, the largest chemical shift of +0.68 eV from 27.26 eV in the metal to 27.94 eV is found for the film with $x$=1.16, while a smaller shift of 0.1 eV occur for $x$=0.30 with a similar shift in the composition range $x$=0.15. The binding energy determined for the Zr *4p$_{3/2}$* peak of the hydride with $x$=1.16 is slightly lower than the 28.1 eV and the 28.0 eV values reported in ref. [9] and [12], respectively. For both these studies, the spectra were recorded for oxidized bulk samples, where the interaction of Zr atoms with O causes the Zr *4p* peak to shift to a higher binding energy. Finally, the *4p$_{3/2}$* peak position of the metal reference of 27.26 eV is in agreement with the reported value of 27.1 eV in ref. [26]

The observed binding energies of the α-Zr spin-orbit split *4p$_{3/2,1/2}$* levels are 27.26 eV and 28.60 eV (1.3 eV splitting). This is consistent with tabulated values of 27.1 eV and 28.5 eV [26]. At the shallow Zr *4p* level, we observe both a significant broadening of the spectra and a high-energy shift in the binding energy relative to metallic α-Zr. Only for the film with $x$=0.30 some oxygen is left even after sputtering. This was also observed by time-of-flight energy elastic recoil detection analysis (ToF-ERDA) showing a slightly higher oxygen content of 0.7% for this film [14] compared to 0.2% for the films with $x$=0.15 and $x$=1.16.

Figure 4 shows a set of high-resolution VBs XPS spectra of the ZrH$_x$ thin films (0.15, 0.30 and 1.16) in comparison to α-Zr metal. In the region 0-2 eV from the $E_F$, the spectra are dominated by the Zr *4d* valence states with a *4d$_{5/2-3/2}$* spin-orbit splitting of ~1.3 eV, most clearly observed in pure α-Zr ($x$=0). For $x$=0.15, the shape of the VB-XPS spectrum is very similar to that of pure α-Zr with a larger broadening and appear to be very little affected by the small hydrogen content. On the contrary, for $x$=1.16, a major spectral redistribution of intensity changes of the shape of the Zr *4d* band that is further broadened so that the spin-orbit splitting is no longer resolved. The most prominent new feature appears around 6 eV binding energy and is due to the strong H *1s* – Zr *4d* hybridization and bonding. For *x*=0.30, a significant broadening of the Zr *4d* valence states at ~1eV is observed. The largest change is in the double-peaked feature between 6 and 7.8 eV. This feature is consistent with previous observations in bulk material [23] with $\delta$-ZrH$_{1.52}$, $\delta$-ZrH$_{1.65}$ and $\delta$-ZrH$_{1.74}$ where the height of the Zr *4d* peak close to E$_F$ decreases and the Zr-H bonding peak increases with increasing hydrogen content. These observations suggested a donation of Zr *4d* electrons towards the Zr-H bond, while the Zr-Zr bond weakens. For higher hydrogen content, (*x*=1.94) the states closest to the E$_F$ were enhanced due to a fcc-fct phase transition [9]. However, bulk materials are often hampered by the appearance of superimposed O *2p* states between 5-10 eV i.e., in the same energy region as the Zr-H bonding peak. As ZrO$_2$ is known to have prominent peaks of O *2p* states between 5-10 eV and O *2s* states around 22.5 eV from the E$_F$ [12]. Thus, in bulk materials, the contribution of the Zr-H bond peak cannot be fully distinguished or separated (disentangled) from the O *2p* states due to oxygen impurities.

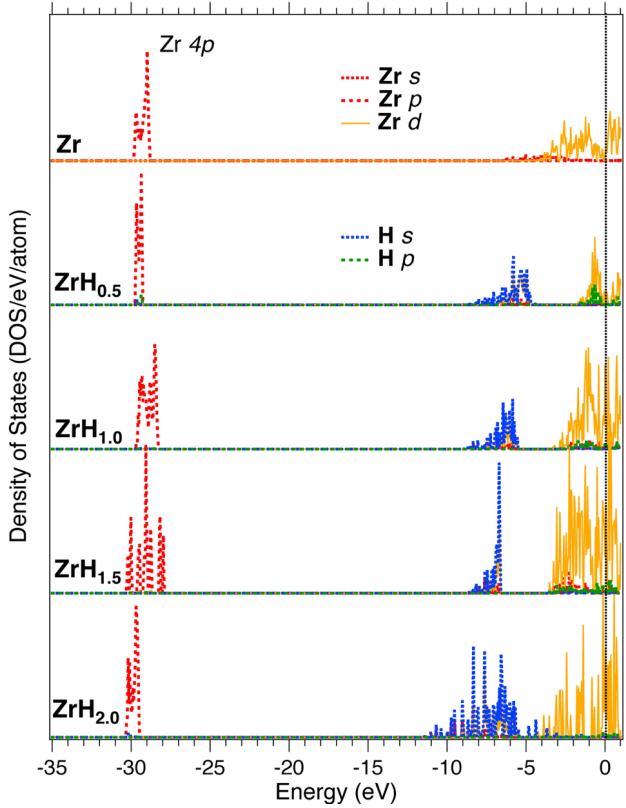

**Figure 5:** Calculated density of states (DOS) for $\alpha$-Zr and $\delta$-ZrH$_x$, where *x*=0.5, 1.0, 1.5 and 2.0.

Figure 5 shows calculated density of states (DOS) for crystalline $\delta$-ZrH$_x$ for *x*=0, 0.5, 1.0, 1.5 and 2.0. As observed, for *x*=0, the states within 4 eV from the E$_F$ are dominated by Zr *4d* states while strong Zr *4p* states occur at 29-30 eV from E$_F$. This is also the case for increasing *x* but a prominent peak in the energy region 5-10 eV is mainly due to H *1s* states that does not exist in pure Zr. At 5-10 eV below E$_F$, bonding H-H *s-s* $\sigma$, Zr-H *d-s* $\sigma$ and Zr-Zr *d-d* $\sigma$-interactions occur while antibonding H-H *s-s* $\sigma*$ interactions appear above E$_F$. We find that the H *1s*-character in $\delta$-ZrH$_x$ strongly depends on the hydrogen content and increases with *x*. Notably, a deep DOS minimum at 2-4 eV occurs for *x*=0.5 when hydrogen is introduced. Closer to the E$_F$, Zr *4d*-$e_g$ states dominate at 0-2 eV while H *1s*-states are negligible. For $\delta$-ZrH$_2$, there is an intense peak at the E$_F$ of Zr *4d*-$t_{2g}$ character that dominate and cause instability of the crystal structure at 0 K. At finite temperature, phonon interactions and tetragonal (Jahn-Teller) distortion of the structure leads to a splitting of the intense $t_{2g}$ peak at the $E_F$ and a pseudogap develops. This leads to lowering of the total energy of the system and restabilization of the structure. Thus, the $t_{2g}$ states at E$_F$ are reduced at finite temperatures and likely reduces the conductivity and other transport properties. This is consistent with previous DFT studies including the $\delta$- and $\varepsilon$-phases [29] [30] [31] [32] of ZrH$_x$ while comparisons to experimental electronic structure studies of the Zr-H bonds of thin films have been lacking. Thus, the order and bonding of the hydrogen atoms in the octahedral sites has not been well understood.

Figure 6 shows the structures (left) and calculated total energies (right) of ZrH$_x$ for $x$=0.5, 1.0, 1.5 and 2.0 (top to bottom). Here, Jahn-Teller distortion is predicted for high hydrogen content ($x$>1), where the cubic δ-structure becomes unstable and lowers the symmetry. On the contrary, for low hydrogen content, the cubic structure is stable ($x$=0.5) or metastable ($x$=1.0). For $x$=2.0, the minimum of $c/a$ is located at 0.886, for $x$=1.5 at 0.868 while for $x$=1 there is a metastable state for $c/a$=1.0 and a lowest stable state for $c/a$=0.792. Thus, the cubic δ-ZrH$_x$ phase is metastable as thin film up to $x$=1.16, while for higher H-contents, the ZrH$_x$ structure is predicted to be distorted by the Jahn-Teller effect.

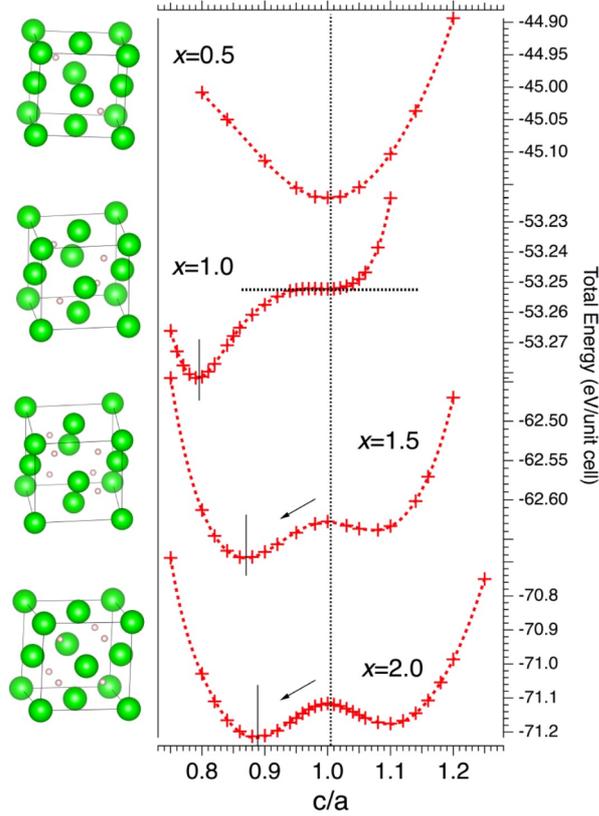

**Figure 6** (Left panels): Model structures. (Right panel): Calculated total energies for ZrH$_x$, where $x$=0.5, 1.0, 1.5 and 2.0.

To evaluate the charge-transfer dependence on $x$, we calculated the BADER charges that are listed in Table I together with the lattice parameters for different compositions. Unit cells of δ-ZrH$_x$ (Fig. 6, left panels), where the H-atoms where replaced by vacancies were used as model systems for different $x$. Among the hydrides, the structure and stability properties of ZrH$_2$ have been of particular interest where the attention has been focused on the δ-ε phase transition induced by stress, and the influence of the Jahn-Teller distortion to stabilize the structure [30] [33]. As the electronegativity of Zr (1.33) is lower than for H (2.2), the formal oxidation states of H and Zr in ZrH$_2$ are -1 and +2, respectively. Generally, the predicted charge-transfer from Zr increases with increasing $x$ and lattice parameter. In comparison to the pure Zr and H elements, the Bader charge of on the Zr atoms in e.g., ZrH$_2$ increases significantly by +1.49$e$, i.e., significant charge is withdrawn from the $4d^2 5s^2$ valence orbitals of Zr. According to the Bader analysis of ZrH$_2$, the charge is transferred to the lowest unoccupied orbital, of the hydrogen atoms ($1s^1 \rightarrow 1s^{1.75}$), where the Bader charge decreases by -0.75$e$ on each atom. A distortion of the cubic δ-ZrH$_x$ structure causes a slight reduction of the charge transfer.

The combined *3d* and *4p* core-level and *4d* valence band studies show several interesting effects. For the hydride system, we observe a significant reduction of the Zr *4d* band within ~3 eV from the E$_F$ influencing the shallow *4p* core level by a broadening and an asymmetric redistribution of intensity towards higher binding energy. There is no trace of oxygen on the samples, except for $x$=0.30 where small signs of O *2p* and O *2s* intensity can be distinguished. For the hydrides, a bonding band at 6 eV below the E$_F$ does not occur in pure Zr and is related to strong H *1s* – Zr *4d* hybridization. This is also accompanied by a chemical shift with binding energies that are 0.6 eV higher than pure α-Zr indicating significant charge-transfer towards H, in particular for $x$=1.16. In our x-ray absorption study [34], a chemical shift towards higher energies in comparison to Zr metal was found to be due to changes in the oxidation state that depends on the structure and the formation of Zr-H bonding at sufficient hydrogen loading. Previous valence band XPS experiments have shown a dominant Zr *4d* peak at 1 eV and another peak around 6.4 eV corresponding to a Zr-H bond [35], consistent with DOS calculations [31] [29] [30]. The chemical shift of the Zr edges towards higher bonding energies confirms a significant charge-transfer from Zr towards H with increasing hydrogen content. It has been argued that a tetragonal distortion of the cubic δ-ZrH$_x$ structure in combination with phonons make the structure stable [30]. However, as shown in Fig. 6, this is only the case for high H-content, while at low H-content, the cubic structure is stable or metastable and does not distort. Thus, for low H-content, the cubic δ-structure with tetrahedral Zr sites favourable affects the materials properties in comparison to a tentative octahedral coordination that for hydrogen is inhibited according to the Hägg rules on size, coordination number and electronegativity.

For ZrH$_x$ films deposited by rDCMS, we have previously shown that it is possible to grow films with tailored compositions by altering the concentration of H$_2$ in the plasma [14]. The properties of the films resemble those of the parent Zr metal where increasing hydrogen content in the films yields higher resistivity seen from the values ranging from ~70 to ~120 μΩ cm in comparison to 42.6 μΩ cm for α-Zr [36]. The measured hardness values are around 5.5 GPa, which is harder than the ~3 GPa determined for bulk α-Zr, using nanoindentation [37].

In studies of bulk ZrH$_x$, there have been problems with oxide as Zr is known to slowly form stable ZrO$_2$ on the surface and at the grain boundaries [12]. Previous studies have thus been limited to more or less oxidized polycrystalline bulk materials that contain randomly ordered crystallites with grain boundaries that affect the results of electronic structure characterization. Moreover, investigations of the electronic structures of understoichiometric transition metal hydrides have been scarce. Thin films synthesized by reactive magnetron sputtering offer an alternative route to grow metal hydrides such as ZrH$_x$ with well-defined properties and at a fraction of time (deposition time of about 2 min for an 800nm film) compared to traditional hydration/perculation processes that normally takes 1-2 days or weeks depending on the H-content [7]. A further advantage of the films is the low level of contaminants, in particular, the oxygen content that is of the range of 0.2-0.7% as determined by ToF-ERDA [14]. Although *3d* core-level shifts in ZrH$_x$ [14] has previously been observed, indicating charge-transfer from Zr to H, a deeper analysis including branching ratio and singularity index is necessary for the understanding how the electronic properties influence the conductivity. For low H-content ($x$=0.15 and 0.30), the films contain a mixture of metallic α-Zr and δ-ZrH$_x$ phases and the metallicity remains. For higher H-content ($x$=1.16), a homogeneous oriented, textured and understoichiometric δ-ZrH$_x$ phase if formed with a lower singularity index and smaller branching ratio that overall indicate more ceramic properties than for the lower H-contents. Producing highly stoichiometric and phase pure materials are thus important for optimizing electrical and hardness properties in a vast number of applications.

## Conclusions

Early transition metal hydrides are a new class of thin film materials where the cubic δ-ZrH$_x$ phase is metastable up to $x$=1.16, while for higher H-contents, the ZrH$_x$ structure is predicted to be distorted by the Jahn-Teller effect both in thin films and bulk synthesized materials. By combining valence-band and shallow core-level X-ray photoelectron spectroscopy with electronic structure calculations, we have investigated the electronic structure and chemical bonding of ZrH$_x$ thin films with $x$=0.15, 0.30, and 1.16 in comparison to α-Zr metal. For the understoichiometric δ-ZrH$_{1.16}$ film, there are significant chemical shifts of 0.68 and 0.51 eV for the *4p$_{3/2}$* and *3d$_{5/2}$* peak positions towards higher energies compared to the metal values of 27.26 and 178.87 eV. We find that even though there is a significant charge-transfer from the Zr *4d* states towards the H *1s* states in δ-ZrH$_{1.16}$, there are still states crossing the Fermi level that signifies metallicity. This is due to the fact that there is a significant asymmetric redistribution of spectral intensity from the shallow Zr *4p* core-level towards the Zr *4d* valence band that accompany the charge-transfer. There are important changes at the Fermi level with a splitting of the *4d* band into a pseudo gap where the Fermi level is located in a local minimum that stabilize the structure and minimize the total energy of the system. For the hydrides, a bonding band around 6 eV below the Fermi level does not occur in pure Zr and is related to strong H *s* – Zr *4d* hybridization. This is also accompanied by a chemical shift with binding energies that are 0.68 eV higher than pure α-Zr indicating significant charge-transfer towards H, in particular for *x*=1.16.

## Acknowledgments
The research leading to these results has received funding from the Swedish Government Strategic Research Area in Materials Science on Functional Materials at Linköping University (Faculty Grant SFO-Mat-LiU No. 2009-00971). MM acknowledges financial support from the Swedish Energy Research (no. 43606-1), the Swedish Foundation for Strategic Research (SSF) (no. RMA11-0029) through the synergy grant FUNCASE and the Carl Trygger Foundation (CTS16:303, CTS14:310). Dr. Grzegorz Greczynski is acknowledged for assistance with the measurements of the Zr 3d core-level XPS spectra.